\documentclass{npw}
\usepackage{graphicx}

\begin{document}

\title{Effect of a realistic three-body force on the spectra of medium-mass hypernuclei}
\author{
   P. Vesel\'{y}\email{p.vesely@ujf.cas.cz} \\
  \it Nuclear Physics Institute, Czech Academy of Sciences, 250 68 \v{R}e\v{z}, Czech Republic \\
 G. De Gregorio \\
  \it Nuclear Physics Institute, Czech Academy of Sciences, 250 68 \v{R}e\v{z}, Czech Republic \\
  \it INFN Sezione di Napoli, Naples, Italy \\  
  J. Pokorn\'{y} \\
  \it Faculty of Nuclear Sciences and Physical Engineering, \\ 
  \it Czech Technical University in Prague, 115 19 Prague, Czech
Republic \\ 
  \it Nuclear Physics Institute, Czech Academy of Sciences, 250 68 \v{R}e\v{z}, Czech Republic  
}


\date{26 September 2018}

\maketitle

\begin{abstract}

We adopt the Hartree-Fock (HF) method in the proton-neutron-$\Lambda$ (p-n-$\Lambda$) formalism and the nucleon-$\Lambda$ Tamm-Dancoff Approximation (N$\Lambda$ TDA)  to study the energy spectra of medium-mass hypernuclei. The formalism is developed for a potential derived from effective field theories which includes explicitly the 3-body $NNN$ forces plus the $YN$ LO potential.   The energy spectra of selected medium-mass hypernuclei are presented and their properties discussed. The present calculation is the first step of a project devoted to {\it ab initio} studies of hypernuclei in medium and heavy mass regions. This may   provide a guide for a better understanding of the $YN$ interactions at momentum scales not accessible in few-body hypernuclei.

\end{abstract}
  

\section{Introduction}
\label{Sect.01}

Hypernuclei have been studied within many theoretical approaches. Let us mention some of them. Faddeev or Yakubovsky equations based on realistic baryon-baryon interactions have been used to study the     $A=3$ and $A=4$ systems \cite{Nogga}. {\it Ab initio}   calculations of spectra of light hypernuclei up to the $p$-shell have been performed within no-core shell model \cite{Wirth1,Wirth2} and fermionic molecular dynamics \cite{Neff}. Medium-mass and heavy hypernuclei have been studied mostly within phenomenological models. The list includes calculations performed within the antisymmetrized molecular dynamics \cite{Isaka}, the shell model \cite{SM}, relativistic  \cite{Mares} and Skyrme   \cite{Rayet,Schulze} mean field theories. 

There are very few theoretical approaches which use realistic interaction to describe the hypernuclei above the $p$-shell. Among them, let us mention the auxiliary field diffusion Monte Carlo (AFDMC) method \cite{Lonardoni1,Lonardoni2} which adopts local baryonic forces and the hypernuclear mean-field model using realistic nucleon-nucleon ($NN$) and nucleon-$\Lambda$ ($N\Lambda$) interactions   \cite{Ves1,Ves2}. 

 The first requirement for a mean field approach is that the calculation must provide a satisfactory description   of the density distributions, root-mean-square (rms) charge radii, and single-particle energies of the nucleons in the nuclear cores. In fact, the single-particle energies of $\Lambda$ in hypernuclei are mainly affected by the nuclear core rms radii and the density distributions. This is a consequence of the Bertlmann-Martin inequalities which link the splitting of the $\Lambda$ $0s$ and $0p$ single-particle levels with the nuclear rms radii \cite{Bertl}. On the other hand,  HF  is not able to reproduce the correct nuclear binding energies.  Ground-state correlation are to be included for this purpose \cite{gs}.  

The HF calculations using only realistic 2-nucleon $NN$ interactions underestimate the nuclear radii \cite{Hergert} and overestimate the relative gaps among the nucleon single-particle levels \cite{Bianco}.   Their correct description was partly achieved by including a phenomenological repulsive density dependent  (DD) term simulating the $NNN$ interaction \cite{Hergert,Bianco}. In our previous  work \cite{Ves2}, in fact, such a DD force was added to the optimized chiral $NN$ potential NNLO$_{\rm{opt}}$ \cite{Ekstroem1} to generate the hypernuclear mean field. 

In order to avoid free phenomenological parameters it is desirable to use a 3-body $NNN$ interaction deduced from effective field theories. Such a term is included in the chiral $NN + NNN$ NNLO$_{\rm{sat}}$ potential derived recently \cite{Ekstroem2}. This interaction was optimized so as to reproduce the low-energy nucleon-nucleon scattering data as well as binding energies and radii of selected nuclei up to oxygen and carbon isotopes \cite{Ekstroem2}. It was used successfully  in  a recent HF + Random Phase Approximation (RPA) calculation of nuclear multipole resonances \cite{Wu}  and in a calculation based on the self-consistent Green's function approach (SCGF) to study the potential bubble nucleus $^{34}$Si \cite{Duguet}.

In this paper we adopt the above potential within the HF method in the proton-neutron-$\Lambda$ formalism  to generate the mean field for hypernuclei with an even-even nuclear cores. 

Additionally, we develop the  N$\Lambda$ TDA  to describe  hypernuclei  consisting of one $\Lambda$ coupled to the even-odd or odd-even nuclear cores. The 3-nucleon $NNN$ force is included in the derivation of the N$\Lambda$ TDA  only at the 2-body level of its normal-ordered form. The residual 3-body term gives zero contribution and is, therefore, ignored.

In this paper, we first implement the chiral $NN + NNN$ NNLO$_{\rm{sat}}$ potential as force acting among nucleons, and we study the influence of its 3-nucleon $NNN$ term on the rms charge radii, and nucleon single-particle energies of the nuclei $^{16}$O, $^{40}$Ca. 

We investigate  the effect of the $NNN$ force on the single-particle energies of $\Lambda$ in   $^{17}_{\Lambda}$O and $^{41}_{\Lambda}$Ca  and on the energy spectra of  the   $^{16}_{\Lambda}$O, $^{40}_{\Lambda}$K. The interaction of $\Lambda$ with nucleons is described by the $N\Lambda - N\Lambda$ channel of the chiral $YN$ LO potential \cite{Haidenbauer}. 
We do not take into account the influence of the $\Lambda - \Sigma$ mixing. This issue is discussed in the final Section of the text. 

The methods presented here are the first step of our project.  Our long-term goal is to go beyond the mean-field approximation by including more complex many-body configurations.

This goal will be achieved by a straightforward  extension to hypernuclei of the  Equation of Motion Phonon Method (EMPM) developed for nuclear structure studies \cite{EMPM1}  and adopted extensively for light and heavy even-even \cite{EMPM2,EMPM3,EMPM4} and medium mass odd-even nuclei \cite{EMPM5,EMPM6,EMPM7}. 



\section{Theoretical Formalism}
\label{Sect.02}

The adopted Hamiltonian has the structure
\begin{equation}
H = T + V^{NN} + V^{NNN} + V^{N\Lambda} - T_{CM} , 
\label{Ham}
\end{equation}  
where $T$ is the kinetic energy operators of nucleons and   $\Lambda$,  $V^{NN}$, $V^{N\Lambda}$, and $V^{NNN}$ stand for the 2-body $NN$, $N\Lambda$, and the 3-body $NNN$ interactions. The term $T_{CM}$ is the centre-of-mass kinetic operator
\begin{equation}
T_{CM} = \frac{1}{2[(A-1)m_{N}+m_{\Lambda}]} \left( \sum_{a=1}^{A} \vec{p}^2_{a} + 2 \sum_{a<b} \vec{p}_a . \vec{p}_b  \right) ,
\label{TCM}
\end{equation}
where $A$ is the baryon number,  $\vec{p}_a$  the momentum operator of the $a$-th particle, $m_{N} \approx 938$ MeV and $m_{\Lambda} \approx 1116$ MeV are, respectively, the masses of nucleon and $\Lambda$.   

In the second quatization formalism, the Hamiltonian (\ref{Ham}) becomes  
\begin{eqnarray}
\nonumber
H = \sum_{ij} t^{\rm{p}}_{ij} a^{\dagger}_i a_j + \sum_{ij} t^{\rm{n}}_{ij} b^{\dagger}_i b_j + \sum_{ij} t^{\Lambda}_{ij} c^{\dagger}_i c_j \\
\nonumber 
+ \frac{1}{4} \sum_{ijkl} V^{\rm{pp}}_{ijkl} a^{\dagger}_i a^{\dagger}_j a_l a_k + \frac{1}{4} \sum_{ijkl} V^{\rm{nn}}_{ijkl} b^{\dagger}_i b^{\dagger}_j b_l b_k \\ 
\nonumber 
+ \sum_{ijkl} V^{\rm{pn}}_{ijkl} a^{\dagger}_i b^{\dagger}_j b_l a_k +\sum_{ijkl} V^{\rm{p}\Lambda}_{ijkl} a^{\dagger}_i c^{\dagger}_j c_l a_k \\
\nonumber 
+ \sum_{ijkl} V^{\rm{n}\Lambda}_{ijkl} b^{\dagger}_i c^{\dagger}_j c_l b_k \\
\nonumber  
+ \frac{1}{36} \sum_{ijklmn} V^{\rm{ppp}}_{ijklmn} a^{\dagger}_i a^{\dagger}_j a^{\dagger}_k a_n a_m a_l \\
\nonumber  
+ \frac{1}{36} \sum_{ijklmn} V^{\rm{nnn}}_{ijklmn} b^{\dagger}_i b^{\dagger}_j b^{\dagger}_k b_n b_m b_l \\
\nonumber  
+ \frac{1}{4} \sum_{ijklmn} V^{\rm{ppn}}_{ijklmn} a^{\dagger}_i a^{\dagger}_j b^{\dagger}_k b_n a_m a_l \\
+ \frac{1}{4} \sum_{ijklmn} V^{\rm{pnn}}_{ijklmn} a^{\dagger}_i b^{\dagger}_j b^{\dagger}_k b_n b_m a_l ,
\label{Ham2}
\end{eqnarray}
where  $a_i^{\dagger}$ ($a_i$), $b_i^{\dagger}$ ($b_i$), and $c_i^{\dagger}$ ($c_i$)  are, respectively,   the creation (annihilation) operators for protons, neutrons, and $\Lambda$.
The hypernuclear mean field is generated by solving the self-consistent HF equations  
\begin{eqnarray}
\nonumber
t^{\rm{p}}_{ij} + \sum_{kl} V^{\rm{pp}}_{ikjl} \rho^{\rm{p}}_{lk} + \sum_{kl} V^{\rm{pn}}_{ikjl} \rho^{\rm{n}}_{lk} + \sum_{kl} V^{\rm{p}\Lambda}_{ikjl} \rho^{\Lambda}_{lk} \\ 
\nonumber
+ \frac{1}{2}\sum_{klmn} V^{\rm{ppp}}_{ikljmn} \rho^{\rm{p}}_{mk}\rho^{\rm{p}}_{nl} + \frac{1}{2}\sum_{klmn} V^{\rm{pnn}}_{ikljmn} \rho^{\rm{n}}_{mk}\rho^{\rm{n}}_{nl} \\ 
+ \sum_{klmn} V^{\rm{ppn}}_{ikljmn} \rho^{\rm{p}}_{mk}\rho^{\rm{n}}_{nl} = \varepsilon^{\rm{p}}_i \delta_{ij} , \label{HFeqsp} \\
\nonumber
t^{\rm{n}}_{ij} + \sum_{kl} V^{\rm{nn}}_{ikjl} \rho^{\rm{n}}_{lk} + \sum_{kl} V^{\rm{pn}}_{kilj} \rho^{\rm{p}}_{lk} + \sum_{kl} V^{\rm{n}\Lambda}_{ikjl} \rho^{\Lambda}_{lk} \\ 
\nonumber
+ \frac{1}{2}\sum_{klmn} V^{\rm{nnn}}_{ikljmn} \rho^{\rm{n}}_{mk}\rho^{\rm{n}}_{nl} + \frac{1}{2}\sum_{klmn} V^{\rm{ppn}}_{klimnj} \rho^{\rm{p}}_{mk}\rho^{\rm{p}}_{nl} \\ 
+ \sum_{klmn} V^{\rm{pnn}}_{klimnj} \rho^{\rm{p}}_{mk}\rho^{\rm{n}}_{nl} = \varepsilon^{\rm{n}}_i \delta_{ij} , \label{HFeqsn} \\
t^{\Lambda}_{ij} + \sum_{kl} V^{\rm{p}\Lambda}_{kilj} \rho^{\rm{p}}_{lk} + \sum_{kl} V^{\rm{n}\Lambda}_{kilj} \rho^{\rm{n}}_{lk} = \varepsilon^{\Lambda}_i \delta_{ij},
\label{HFeqsL}
\end{eqnarray}
where $\rho^{\rm{p}}_{nm}=\langle \rm{HF}| \textit{$a^{\dagger}_m$} \textit{$a_n$} |\rm{HF}\rangle$, $\rho^{\rm{n}}_{nm}=\langle\rm{HF}| \textit{$b^{\dagger}_m$} \textit{$b_n$} |\rm{HF}\rangle$ and $\rho^{\Lambda}_{nm}=\langle\rm{HF}| \textit{$c^{\dagger}_m$} \textit{$c_n$} |\rm{HF}\rangle$ are the proton, neutron and $\Lambda$ densities, respectively, and  $\varepsilon^{\rm{p}}_i$, $\varepsilon^{\rm{n}}_i$, $\varepsilon^{\Lambda}_i$  the single particle energies.

The solution of the HF equations yields the single-particle energies of $\Lambda$ bound in the even-even nuclear core.


The states of the $\Lambda$ bound in the odd-even or even-odd core are determined in  the N$\Lambda$ TDA.
We first solve the HF Eqs. (\ref{HFeqsp}),  (\ref{HFeqsn}) for the nuclear core 
and, subsequently, solve  Eq. (\ref{HFeqsL}) to obtain the self-consistent single-particle energies $\varepsilon^{\Lambda}_i$ and basis states for $\Lambda$.  We, then, introduce the proton-$\Lambda$ and neutron-$\Lambda$ phonon operators  
\begin{eqnarray}
R^{\dagger}_{\mu,\rm{p}\Lambda} = \sum_{ph} r^{\mu,\rm{p}\Lambda}_{ph} c^{\dagger}_p a_{\bar{h}}, \\
R^{\dagger}_{\mu,\rm{n}\Lambda} = \sum_{ph} r^{\mu,\rm{n}\Lambda}_{ph} c^{\dagger}_p b_{\bar{h}}, 
\label{R-phonon}
\end{eqnarray} 
where $a_{\bar{h}} = a_{n_{\bar{h}} l_{\bar{h}} j_{\bar{h}} m_{\bar{h}}} = (-1)^{j_h + m_h} a_{n_h l_h j_h -m_h}$.
These operators represent a general superposition of the $\Lambda$-particle (p) nucleon-hole (h) configurations. The coefficients $r^{\mu,\rm{p}\Lambda}_{ph}$, $r^{\mu,\rm{n}\Lambda}_{ph}$ are fixed by solving the following eigenvalue equations
\begin{eqnarray}
\nonumber
\sum_{ph} \left( (\varepsilon^{\Lambda}_p - \varepsilon^{\rm{p}}_h)\delta_{pp'} \delta_{hh'} - V^{\rm{p}\Lambda}_{\bar{h}p'\bar{h}'p} \right) r^{\mu,\rm{p}\Lambda}_{ph} = \hspace{1.0cm} \\ 
= (E^{\rm{p}\Lambda}_{\nu} - E_{\rm{HF}}) r^{\mu,\rm{p}\Lambda}_{p'h'} ,  \label{pLTDA}
\\
\nonumber
\sum_{ph} \left( (\varepsilon^{\Lambda}_p - \varepsilon^{\rm{n}}_h)\delta_{pp'}\delta_{hh'} - V^{\rm{n}\Lambda}_{\bar{h}p'\bar{h}'p} \right) r^{\mu,\rm{n}\Lambda}_{ph} = \hspace{1.0cm} \\ 
= (E^{\rm{n}\Lambda}_{\nu} - E_{\rm{HF}}) r^{\mu,\rm{n}\Lambda}_{p'h'} . \label{nLTDA}
\end{eqnarray}


\section{Calculation Details and Results}
\label{Sect.03}
In order to solve the HF eigenvalue problem we represent the Hamiltonian (\ref{Ham2}) in the harmonic oscillator (HO) basis. 

The nucleon and $\Lambda$ HO states depend on the oscillator lengths $b_{N}$ and $b_{\Lambda}$ which are fixed by the frequency $\omega_{\rm{HO}}$ through the relation
\begin{equation}
b_{N(\Lambda)} = \sqrt{\frac{\hbar}{m_{N(\Lambda)}   \omega_{\rm{HO}}}} .
\label{length}
\end{equation}
The HF solutions do not depend on the parameter $\hbar \omega_{\rm{HO}}$ if the basis is large enough. We have put  $\hbar \omega_{\rm{HO}} = 16$ MeV. As we shall see, we reach a good convergence by including  all two-body  and  three-body matrix elements under the restrictions  $\left\{ |ij\rangle : 2n_i + l_i + 2n_j + l_j \leq N_{\rm{max}}  \right\}$ and  $\left\{ |ijk\rangle : 2n_i + l_i + 2n_j + l_j + 2n_k + l_k \leq N_{\rm{max}}\right\}$, with $N_{\rm{max}} = 12$.


\subsection{HF investigation of  $^{16}$O and  $^{40}$Ca cores}
\label{Subsect.03.1}
\begin{figure}[ht]
\includegraphics[width=0.95\columnwidth]{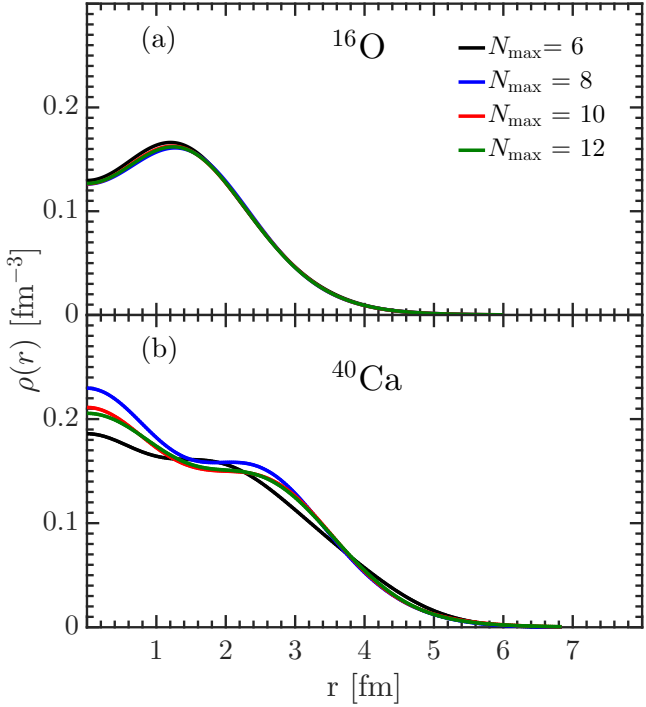}

\caption{Radial density distributions $\rho$(r) of $^{16}$O (a) and $^{40}$Ca (b)  for different $N_{\rm{max}}$.
\label{fig.01}} 
\end{figure} 

\begin{figure}[ht]
\includegraphics[width=0.95\columnwidth]{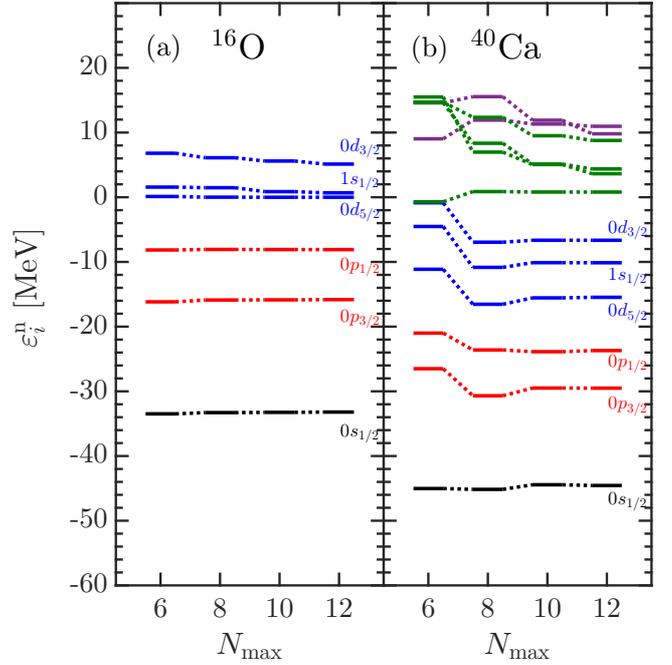}

\caption{The neutron single-particle energies $\varepsilon^{\rm{n}}_i$ of $^{16}$O (a) and $^{40}$Ca (b) calculated for different $N_{\rm{max}}$.
\label{fig.02}} 
\end{figure} 

\begin{figure}[h!]
\includegraphics[width=0.95\columnwidth]{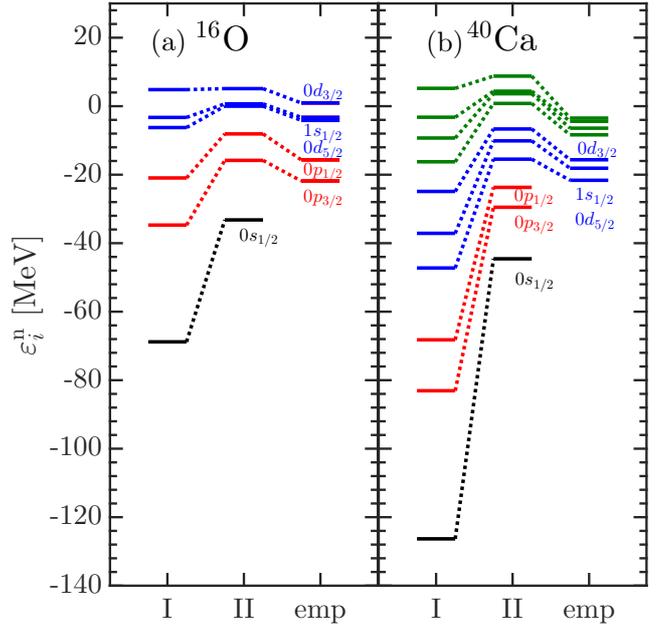}

\caption{The neutron single-particle energies $\varepsilon^{\rm{n}}_{i}$ of $^{16}$O (a) and $^{40}$Ca (b) calculated with $NN$ (I) and $NN + NNN$ (II) interactions. The empirical data (emp) \cite{expsp} are shown for comparison.   
\label{fig.03}} 
\end{figure}  
The radial density distributions, charge radii, nucleon single-particle energies, and binding energies of $^{16}$O, $^{40}$Ca are calculated by solving the HF Eqs. (\ref{HFeqsp}), (\ref{HFeqsn}) for nucleons only ($\rho^{\Lambda} = 0$). 


As shown in Fig. \ref{fig.01}  the radial density distribution in $^{40}$Ca converges with   $N_{\rm{max}}$ more slowly than in $^{16}$O.   In both nuclei, however,  a saturation value is reached.
A good convergence is reached also for the neutron single-particle energies $\varepsilon^{\rm{n}}_i$ (Fig. \ref{fig.02}).




In order to study the effect of the 3-nucleon NNN component of the NNLO$_{\rm{sat}}$ potential on the rms charge radii and neutron single-particle energies we perform calculations with and without this term.



The mean square charged radius is given by
\begin{equation}
<r^2_{\rm{ch}}> = \left(1 - \frac{1}{A}\right) \frac{1}{Z} <r^2_{\rm{p}}> + R^2_{\rm{p}} + \frac{N}{Z}R^2_{\rm{n}} + \frac{3\hbar^2}{4m^2_{\rm{p}}c^2} , 
\label{rch}
\end{equation}
where $<r^2_{\rm{p}}> = \int \rm{d}r \ r^4 \rho_{\rm{p}}(r)$ is the mean-square proton point radius, $R_{\rm{p}} = 0.8775(51)$ fm, $R^2_{\rm{n}} = -0.1149(27)$ fm$^2$, and $\frac{3\hbar^2}{4m^2_{\rm{p}}c^2} = 0.033$ fm$^2$ \cite{Ekstroem2}.

\begin{table}[h!]
\caption{The charge radii  $r_{\rm{ch}} = \sqrt{<r^2_{\rm{ch}}>}$ [fm] of   $^{16}$O and $^{40}$Ca calculated with $NN$ and $NN+NNN$ forces are compared with the experimental data (exp) \cite{Angeli}.}
\begin{center}
\label{tab.01}
\begin{tabular}{cccc}
\hline
$^{A}X$  & $NN$ & $NN+NNN$ & exp \\
\hline
$^{16}$O & 2.19 & 2.77 & 2.70  \\
$^{40}$Ca & 2.58 & 3.54 & 3.48 \\
\hline
\end{tabular}
\end{center}
\end{table}

\begin{table}[h!]
\caption{Binding energies per nucleon $BE/A$ [MeV]
calculated with $NN$ and
with $NN + NNN$ forces in
$^{16}$O and $^{40}$Ca compared to the experimental
values (exp).}
\begin{center}
\label{tab.02}
\begin{tabular}{cccc}
\hline
$^{A}X$  & $NN$ & $NN+NNN$ & exp \\
\hline
$^{16}$O & 7.36 & 2.66 & 7.98  \\
$^{40}$Ca & 11.65 & 2.31 & 8.55 \\
\hline
\end{tabular}
\end{center}
\end{table}

As shown in Table \ref{tab.01}, the charge radii $r_{\rm{ch}}$ calculated with the 2-nucleon $NN$ interaction only are too small. They are enhanced, in much better agreement with the experiments, once  the $NNN$ force is included.   It is worth to notice that   the phenomenological DD term adopted in our previous study produces a similar effect \cite{Ves2}. 

The 3-body $NNN$ force has also the very important effect of reducing the separation between the neutron single-particle levels, as clearly illustrated in Fig. \ref{fig.03} for $^{16}$O, and $^{40}$Ca. An analogous effect was produced by the phenomenological DD term \cite{Ves2}.  
 
The effect of the 3-body $NNN$ force on nuclear radii and nucleon single-particle energies  is caused by its repulsive character. This term, in fact, reduces strongly the binding energies of $^{16}$O, and $^{40}$Ca  which are underestimated by a factor $\sim 3$ and $\sim 3.7$ respectively (Table \ref{tab.02}), indicating that ground state correlations are needed. 


The strong impact on the physical observables induced by more complex configurations emerges from the calculations using  the NNLO$_{\rm{sat}}$ potential  within a Coupled Cluster (CC) \cite{Ekstroem2} and a SCGF approach \cite{Duguet}. 

The effect of the correlations has been also studied within the EMPM using the NNLO$_{\rm{opt}}$ potential \cite{gs}. 




This calculation yielded a contribution to the binding energy coming from two-phonon correlations comparable to the HF contribution. The impact on the charge radius was very modest instead. 

On the ground of this EMPM calculation and the results obtained in Refs. \cite{Ekstroem2,Duguet} with NNLO$_{\rm{sat}}$, we expect that 
the ground state correlations estimated within the EMPM should counterbalance the contribution coming 
from the strongly repulsive $NNN$ part of the NNLO$_{\rm{sat}}$ potential thereby filling to a large extent the gap with experiments.

\subsection{Spectra of $^{16}_{\Lambda}$O, $^{17}_{\Lambda}$O, $^{40}_{\Lambda}$K and $^{41}_{\Lambda}$Ca  hypernuclei}
\label{Subsect.03.2}
\begin{figure}[h!]
\includegraphics[width=0.95\columnwidth]{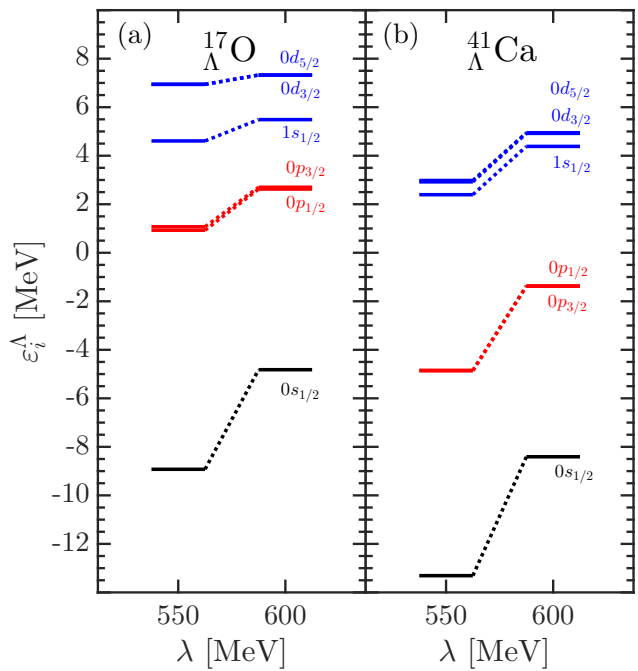}

\caption{The $\Lambda$ single-particle energies $\varepsilon^{\Lambda}_{i}$ in $^{17}_{\Lambda}$O (a) and $^{41}_{\Lambda}$Ca (b) calculated for two different values of the regulator cutoff $\lambda$ of the LO $YN$ potential.
\label{fig.04}} 
\end{figure} 

\begin{figure}[h!]
\includegraphics[width=0.95\columnwidth]{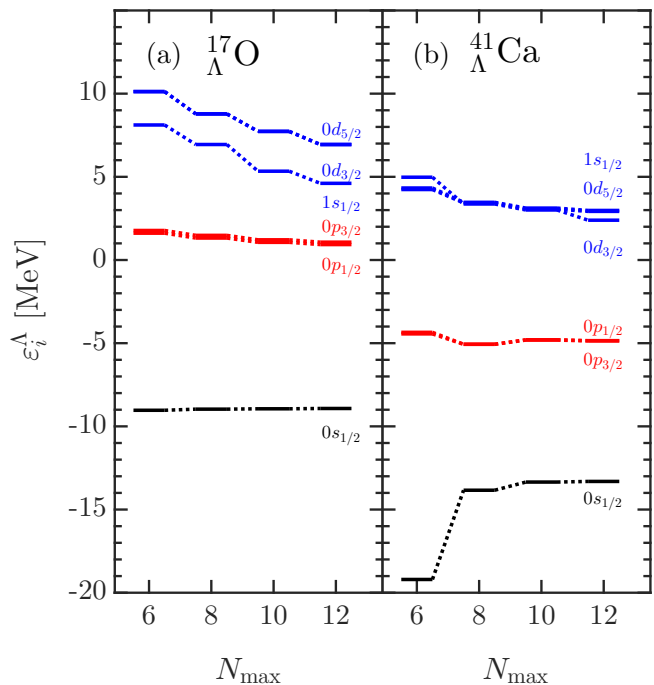}

\caption{The $\Lambda$ single-particle energies $\varepsilon^{\Lambda}_{i}$ in $^{17}_{\Lambda}$O (a) and  $^{41}_{\Lambda}$Ca (b) calculated for different $N_{\rm{max}}$. 
\label{fig.05}} 
\end{figure} 

\begin{figure}[h!]
\includegraphics[width=0.95\columnwidth]{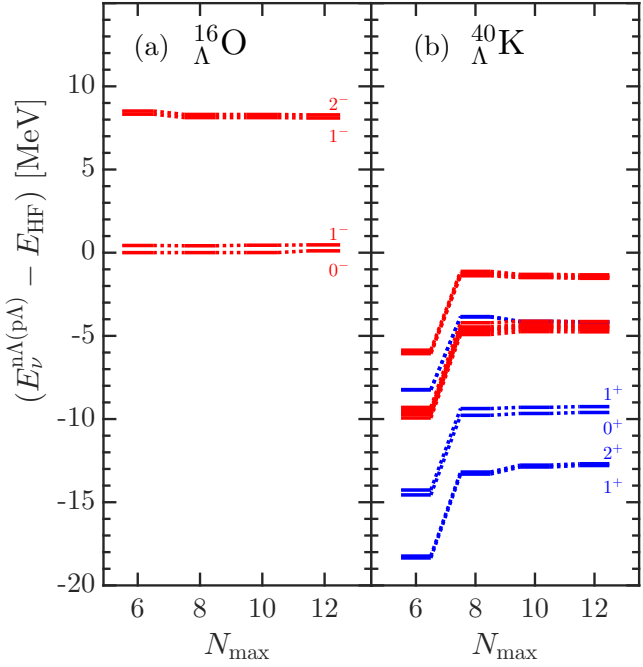}

\caption{The energies $(E^{\rm{n}\Lambda}_{\nu} - E_{\rm{HF}})$ of $^{16}_{\Lambda}$O (a) and $(E^{\rm{p}\Lambda}_{\nu} - E_{\rm{HF}})$ of $^{40}_{\Lambda}$K (b) calculated for different $N_{\rm{max}}$. The red (blue) lines represent the states with the negative (positive) parity. 
\label{fig.06}} 
\end{figure} 
\begin{figure}[h!]
\includegraphics[width=0.95\columnwidth]{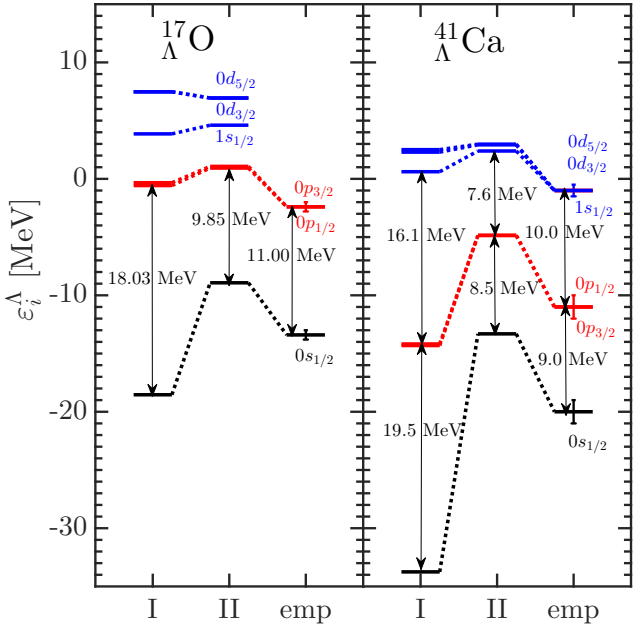}
\caption{The $\Lambda$ single-particle energies $\varepsilon^{\Lambda}_{i}$ in $^{17}_{\Lambda}$O (a) and in $^{41}_{\Lambda}$Ca (b) calculated with $NN$ (I) and $NN + NNN$ (II) interactions. The empirical energies (emp) interpreted from the experimental measurements of $^{16}_{\Lambda}$O \cite{expO} and $^{40}_{\Lambda}$Ca \cite{expCa} are shown for comparison.   
\label{fig.07}} 
\end{figure}

The hypernuclear energy spectra are affected by the strong dependence of the $YN$ LO potential  on the regulator cutoff parameter $\lambda$. We put $\lambda = 550$ MeV. This value yields a more bound $\Lambda$ particle with the energies closer to the empirical values. In general, bigger cutoff $\lambda$ causes an overall upward shift of  the energy levels
(Fig. \ref{fig.04}).  However, the separations among single-particle levels remain relatively stable.



Figs. \ref{fig.05} and \ref{fig.06} point out the good convergence of the energy spectra with respect to the parameter  $N_{\rm{max}}$ of the studied hypernuclei.

As shown in Fig. \ref{fig.07},  the energy gaps among the $\Lambda$ single-particle levels in $^{17}_{\Lambda}$O and $^{41}_{\Lambda}$Ca are significantly reduced  once the $NNN$ interaction is included. The relative distances among $s-$, $p-$, and $sd-$ major shells are in fair agreement with  the empirical levels. The whole spectra, however, are shifted upward  with respect to the empirical one due to the underestimation of the binding energies. Clearly, more complex configurations need to be included.

\begin{figure}[h!]
\includegraphics[width=0.95\columnwidth]{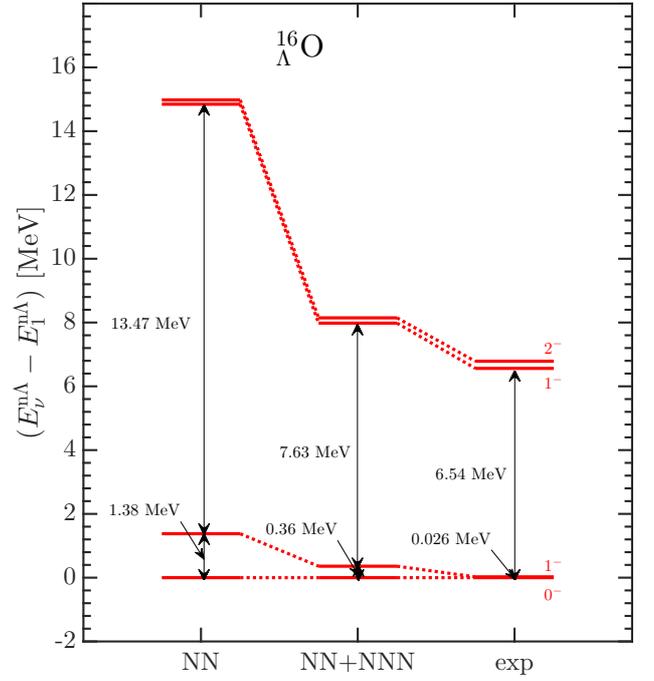}

\caption{The relative energies of $^{16}_{\Lambda}$O calculated with $NN$ and $NN + NNN$ forces. The experimental data (exp) \cite{Tamura} are shown for comparison.    
\label{fig.08}} 
\end{figure}
\begin{figure}[h!]
\includegraphics[width=0.95\columnwidth]{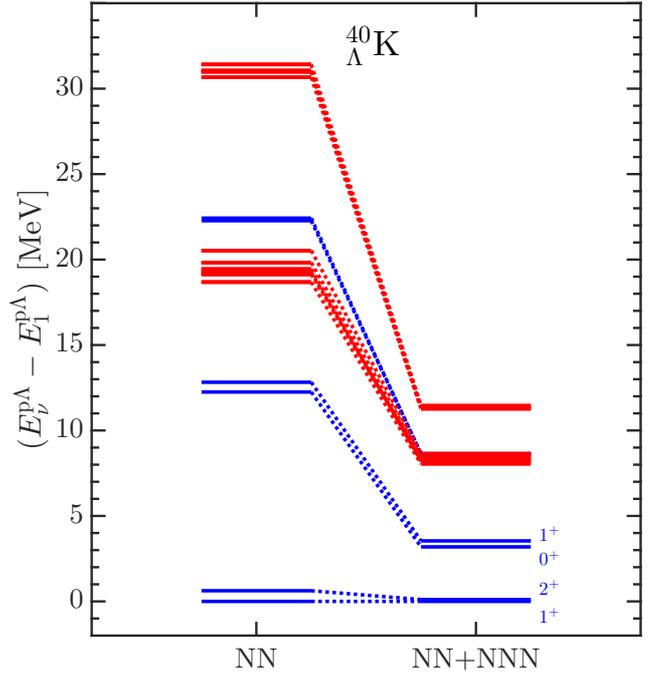}

\caption{The relative energies of $^{40}_{\Lambda}$K calculated with $NN$, and $NN + NNN$ forces. The red (blue) lines represent the states with the negative (positive) parity.
\label{fig.9}} 
\end{figure}
The $NNN$ interaction affects strongly also the relative energy spectra in $^{16}_{\Lambda}$O and $^{40}_{\Lambda}$K.
As illustrated in Fig. \ref{fig.08}, it reduces strongly the gaps among the levels in $^{16}_{\Lambda}$O in very close agreement with the experiments \cite{Tamura} and  an analogous reduction of the separation energies is obtained for $^{40}_{\Lambda}$K (Fig. \ref{fig.9}) where no experimental data are avalaible.




\section{Summary and Outlook}
\label{Sect.04}

The results presented here emphasize the impact of the 3-nucleon $NNN$ component of the chiral NNLO$_{\rm{sat}}$ potential on nuclei and hypernuclei.

In nuclei, in particular $^{16}$O and $^{40}$Ca, the $NNN$ force improves the description of the radial densities by flattening their distributions, enhances the nuclear radii in agreement with the empirical ones, and yields nucleon single-particle energies in close correspondence with the empirical levels. However, its strong repulsive character brings the theoretical binding energies far from the experimental data.

 In hypernuclei, we obtain the $\Lambda$ single-particle levels of  $^{17}_{\Lambda}$O, $^{41}_{\Lambda}$Ca with realistic energy gaps and in close correspondence with the observed levels by an appropriate rigid translation of the whole spectra. This shift is very sensitive to the regulator cutoff $\lambda$ of the $NY$ LO potential. This potential grows linearly with  $\lambda$ and, therefore, affects little the relative energies of the hypernuclear spectra studied here. The relative energies of $^{16}_{\Lambda}$O and $^{40}_{\Lambda}$K also are in better agreement with the available experimental data once the $NNN$ force is included. 

To fill the gap with the experimental nuclear binding energies and to reproduce the absolute energies of hypernuclei, for the adopted value of the regulator cutoff $\lambda$, it is necessary to include more complex excitations. 

We plan to study the effect of the ground-state correlations and the coupling of the $\Lambda$-particle to many particle-hole excitations of the core resorting to the EMPM already mentioned in the introduction.  

This method constructs and solves a set of equations of motion to generate an orthonormal multiphonon basis built of TDA phonons. 
The solution of the eigenvalue equation in the multiphonon basis so constructed yields highly correlated states, including the ground state.
It can be implemented for any realistic Hamiltonian, does not rely on any approximation and takes the Pauli principle into full account.

We extend the method to hypernuclei with even-even and odd-even nuclear cores. In the two cases we couple, respectively, the $\Lambda$ states and the $N\Lambda$ TDA phonons to the multiphonon excitations of the nuclear cores.

We intend to apply this theoretical framework to the hypernuclei $^{40}_{\Lambda}$K, and $^{48}_{\Lambda}$K, whose production is being planned at JLab \cite{JLab}. Our calculations can provide the hypernuclear wave functions needed for the theoretical analysis of this production. 

Another important step is the inclusion of the $\Lambda-\Sigma$ mixing in the $YN$ interaction. This will be accounted for by following the procedure of Ref. \cite{Wirth3} which incorporates the $N\Lambda-N\Sigma$ part of the chiral LO $YN$ interaction into the $N\Lambda-N\Lambda$ channel. The procedure is based on the SRG transformation \cite{SRG} in the Wegner's formulation \cite{Wegner}, which leads to the suppression of the $\Lambda-\Sigma$ conversion terms in the Hamiltonian. 

The SRG transformed interaction must include the 2-body $YN$ as well as the 3-body $YNN$. The SRG transformed $YN$ force alone overestimates by several MeV the binding of $\Lambda$ in the $p-$shell hypernuclei \cite{Wirth3} with respect to the experimental values. 
This over-binding  is reduced to a large extent  by the strongly repulsive SRG induced $YNN$ term. We expect an analogous behavior also in heavier hypernuclei.  In general, it would be also worth to implement other realistic $YN$ interactions like the chiral next-to-leading order (NLO) $YN$ interaction \cite{Haidenbauer2}.

\begin{ack}
We thank Nicola Lo Iudice for his critical reading of the manuscript and helpful comments and suggestions. 
We also thank Carlo Barbieri, Daniel Gazda, and Petr Navr\'{a}til for fruitful discussions, and valuable advices regarding the implementation of the $NNN$, and $YN$ interactions.  
J. Pokorn\'{y} acknowledges Czech Technical University SGS grant No. SGS16/243/OHK4/3T/14. 
Highly appreciated was the access to computing and storage facilities provided by the Meta
Centrum under Program No. LM2010005 and the CERITSC under the program Centre CERIT
Scientific Cloud, part of the Operational Program Research and Development for Innovations,
Register No. CZ.1.05/3.2.00/08.0144. \\ 
\end{ack}


\end{document}